**Electrostatic mechanophores in tuneable light-emitting piezo-polymer nanowires**

*Luana Persano\*, Andrea Camposeo, Aleksandr V. Terentjevs, Fabio Della Sala\*, Eduardo Fabiano, Martina Montinaro, Dario Pisignano\**

Dr. L. Persano, Dr. A. Camposeo, Prof. D. Pisignano
NEST, Istituto Nanoscienze-CNR, Piazza S. Silvestro 12, I-56127 Pisa, Italy
E-mail: luana.persano@nano.cnr.it
Dr. A. Camposeo, Dr. F. Della Sala, Dr. E. Fabiano, Prof. D. Pisignano
Center for Biomolecular Nanotechnologies @UNILE, Istituto Italiano di Tecnologia, via Barsanti, I-73010 Arnesano (LE), Italy
Dr. A. V. Terentjevs
Istituto Nanoscienze-CNR, Via per Arnesano, I-73100, Lecce, Italy
F. Della Sala, E. Fabiano
Institute for Microelectronics and Microsystems (CNR-IMM), Via Monteroni, Campus Unisalento, I-73100 Lecce, Italy
E-mail: fabio.dellasala@le.imm.cnr.it
M. Montinaro, D. Pisignano
Dipartimento di Matematica e Fisica "Ennio De Giorgi", Università del Salento, via Arnesano, I-73100, Lecce, Italy
E-mail: dario.pisignano@unisalento.it

In materials at the solid state, any mechanical action can induce chemico-physical changes which affect the overall or local properties. These variations may involve mass transfer, heating, mechanical stress generation/relaxation, or redistribution of electronic or ionic charges.[1-3] In particular, the capability to control the optical properties of molecules by a mechanical stress can activate or deactivate chemical reactions, and enable new functionalities for monitoring and sensing.[2,4] This is both a fundamental challenge in the broad field of nanosciences and a significant advantage for applications in medical diagnostics, and for signal transduction in bio-interfaces. Mechano-photo conversion has been reported, based on dynamic molecular recognition at the air-water interface.[5-7] Recently, anisotropic compressive stresses at nN-scale have been found to induce photo-physical changes in conjugate polymer nanoparticles[8] and in single organic dye molecules.[9] In another report, uniaxial tensile forces have been applied to polymer foils doped with an oligo-paraphenylenevinylene derivative, leading to a blue-shift of the emission by 1.2 nm.[4] With focus on practical applications, the capability to control electro-optical processes through the modification of a piezoelectric potential, in turn led by an external stress, has been introduced by the Wang's group in 2010.[10] This piezo-phototronic effect is based on the use of a strain-induced piezo-potential at a metal-semiconductor interface, which tunes the charge transport across the junction[11,12] and the quantum efficiency of nanostructured light emitting diodes.[13,14] Piezo-phototronics is thus related to the simultaneous presence of piezoelectricity, emission excitation, and semiconducting behaviour in the involved materials. Seeking for alternative platforms, extending further the control of optical properties through mechanical forces on flexible materials, might not only help to elucidate the fundamental mechanisms of opto-mechanics at nano-scale, but also open new perspectives for the development of next sensing and optoelectronic devices.

In this work, we introduce electromechanical coupling through piezoelectric polymer chains to tune the emission of organic molecules in active nanowires. Our system is made of highly bendable arrays of counter-ion dye-doped nanocomposite wires made of a poly(vinylidenefluoride) (PVDF) copolymer. A reversible red shift of the dye emission is found upon applying dynamic stress during highly accurate bending experiments. For purpose of comparison, no significant shift is found in either organic nanowires without counter-ions or a non-piezoelectric polymer. Using density functional theory (DFT) calculations, we show that these photophysical properties are associated with mechanical stresses applied to electrostatically-interacting molecular systems, namely to counterion-mediated states involving the light-emitting molecules and the positively charged region of the piezoelectric polymer chain. This unveiled, electrostatic class of supramolecular mechanophores, as functional stress-sensitive units,[15-17] might impart new functionalities in anisotropic, hybrid molecular nanosystems.

Poly(vinylidenefluoride-*co*-trifluoroethylene) [P(VDF-TrFe)] directly crystallizes from the melt or solution into the ferroelectric phase analogous to β-PVDF. It is characterized by high stability at room temperature and high achievable degree of crystallinity (>90%).[18] PVDF and its copolymers have been largely investigated, and devices based on their nanofibers have shown enhanced piezoelectric properties, as well as highly flexible design and lightweight construction.[19-21] Here light-emitting, composite nanowires are realized by electrospinning, using a fast rotating disk collector to generate free-standing arrays, and investigated under cyclic and dynamic stress conditions (schematics in **Figure 1**a,b). As reference, different molecular dyes as well as a plastic, non-piezoelectric polymer (polymethymethacrylate, PMMA) are also tested. Prior to electrospinning, polymer solutions are doped with 1-1.5% w/w of light-emitting molecules (Figure 1c,d and Figure S1 in the Supporting Information). Scanning electron microscopy (SEM) images indicate that the nanocomposite P(VDF-TrFe) wires mostly have a cylindrical shape with smooth surfaces and average diameter of 150-180 nm (**Figure 2**a and Figure S2). Photographs of the arrays of nanowires are shown in Figure 2b-e, where the different colours are indicative of the different used dyes, rhodamine 700 (LD700, Figure 2b,d), disodium-1,3,5,7,8-pentamethylpyrromethene-2,6-disulfonate-difluoroborate complex (pyrromethene, Figure 2c) and [2-[2-[4-(dimethylamino)phenyl]ethenyl]-6-methyl-4H- pyran-4-ylidene]-propanedinitrile (DCM, Figure 2e). The obtained morphology of the nanowires is related to the solution concentration and entanglements, and to the properties of the solvent used.[22,23]

Nanocomposite flexible wires are positioned on top of a 76 μm-thick Kapton (PI) film, and devices are built by establishing electrical contacts to the ends of the arrays of fibers (typical length ≅ 1.5 cm). Bending experiments are performed by using two synchronized linear stages, with the edges of the sample fixed with two stripes positioned on top of each stage. A high-speed camera precisely records the bending movement to determine the instantaneous values of the curvature radius and strain rate. The plot of the out-of plane bending amplitude vs time for three bending cycles is reported in Figure S3. Each half-period (from flat to the maximum bent state and return) is well described by the laws of uniformly accelerated motion. The values of the bending radius ($R$) and strain rate ($\dot{\varepsilon}$) at specific instant frames are reported in Figure 2f-j. At the maximum amplitude of bending, $R$ is as low as 4 mm, and the strain rate zeroes. The instantaneous values of the bending radius and of the maximum exerted strain ($\varepsilon_m$) are displayed in Figure S4. During each bending cycle, $\varepsilon_m$ approaches 1.5% and the corresponding tensile stress, $\sigma$, is ~2.5 N/mm$^2$ ($\sigma \cong \varepsilon_m E$, with $E$ Young modulus, Figure S5) Under bending, a periodic alternation of negative and positive voltage output peaks (0.5-1 V) corresponds to the application and release of the buckling stress to P(VDF-TrFe)-based wires (Figure S6).

In addition, during bending cycles, the emission signal of the embedded dyes, photo-excited by a laser with wavelength 405 nm and power 0.2 mW, is acquired avoiding any spurious effect related to movement (spectra shown in Figure S7, experimental details in Supporting Information, Note 1 and in Figures S8-S10). Briefly, the experimental system used to collect the light emitted by the samples is carefully implemented to be synchronous to the mechanical movement and to keep fixed the laser spot on the sample surface. A region of $3 \times 10^{-2}$ mm$^2$ is excited during each bending cycle. The applied strain is of tensile type, since the neutral strain position, where bending-induced compressive and tensile deformations cancel each other, is located within the Kapton layer.[24] The effect of strain application is studied by repetitively recording the emission spectra with an integration time of 20 ms ($<< \dot{\varepsilon}^{-1}$). **Figure 3**a displays the PL spectra of a P(VDF-TrFe)/LD700 array collected during mechanical bending at different times. Data show that the emission of P(VDF-TrFe)/LD700 is reversibly red-shifted, that is optical transitions move towards lower energies, upon tensile strain (Figure 3a). Larger values of the red-shift are recorded at $R \sim$ 5 mm and $\dot{\varepsilon} \sim 0.04$ both during the application ($t$=0.27 s) and release ($t$=0.68 s) of the buckling stress. Such behaviour is cyclic and can be captured by measuring the PL response over periodic bending and unbending movements. In Figure 3b the plot of the emission peak wavelength vs. time (continuous lines) for three different P(VDF-TrFe)/LD700 samples is reported, evidencing a measured red-shift by at least 2-5 nm compared to the emission wavelength measured in static and planar conditions (horizontal dashed lines in Figure 3b). Sample to sample variations (a few per mil) of the shifts are attributable to minor disuniformities in the incorporation of dyes during electrospinning, or possible local aggregates of the emitters, which affect the initial conformation and local potential energy surface (PES) of the molecules. Also, effects of possible metastable states are appreciable through the $\dot{\varepsilon}$-dependence of the optical red-shifts, with maximum wavelength variations corresponding to higher strain rate values and slightly lower variations found at $\dot{\varepsilon} \cong 0$ as shown in Fig. 3a,b ($R \sim$ 4 mm, $t$=0.45 s). Local mechanisms for metastability could involve diffusional reconfiguration of dyes in the polymeric host (up to a few nm within $\sim$0.1 s, for typical diffusion coefficients of $10^{-13}$-$10^{-15}$ cm$^2$/s).[25,26] Several experiments are performed evidencing these results.

In addition, since during bending the vertical distance between the area excited on the sample and the optical collection system varies up to 1 cm, a calibration procedure is implemented to precisely determine the expected spectral shift due to changes of collection angle (orange dotted lines in Figure 3b and Supporting Information). In nanowires which are based on the same piezoelectric matrix but embedding a dye without counter-ions (pyrromethene), no significant modification of the transition energies is found, and only slight fluctuations of the emission peak associated to the optical detection during sample bending are measured (Figure 3c). In case of plastic, non piezo-electric samples, no substantial shift of the emission can be attributed to mechanical action during bending, and the observed spectral variations are always in line with those induced by the detection geometry variations for chromophores both with (LD700) and without (DCM) counter-ions (Figure S11). These results suggest that a synergistic mechanism takes place in the nanocomposites, which is initiated by mechanical action, then being assisted by the electrostatic interaction between counter-ions and the positive region of piezo-dipoles at the sub-nanoscale.

To rationalize these findings, we analyze the P(VDF-TrFe)/LD700 system by first-principles calculations. In the ground-state (zero stress) a given LD700 molecule with its $ClO_4^-$ counter-ion is fully surrounded by the P(VDF-TrFe) polymer chains, due to the low dye density. **Figure 4**a shows the molecular structure, in the case of a single chain, obtained from DFT geometry optimization (computational details in Methods). Fourier-Transform Infrared (FTIR) measurements for light polarized parallel to the length of the composite fibers highlight dye features (at about 1500 and 1600 $cm^{-1}$) which well-compare with computed IR absorption spectra polarized along the long molecular axis for LD700 (Figure S12). This indicates substantial orientation of the dopants parallel to the fibers, namely to the P(VDF-TrFe) backbones. The counter-ion strongly interacts with the positively charged hydrogens of the P(VDF-TrFe) piezo-dipoles (see the electrostatic potential map in Figure 4b) as well as with the LD700 molecules (dotted lines in Figure 4a highlight the shortest distances to $ClO_4^-$).

The interaction is mainly electrostatic. DFT calculations show that the PES of the LD700/ClO$_4^-$ coupled system is rich of local minima, also affected by neighbour chains. When strain is applied, it induces a spatial shift between the polymer chains and the LD700/counter-ion. Hence, the relative orientation between LD700 and ClO$_4^-$ will dynamically change, i.e. the system will likely transit into other local minima of the calculated PES. For instance, Figure 4c shows the absorption spectra of LD700 for the two different positions of the counter-ion (displayed in the inset), as computed from time-dependent DFT. The two configurations correspond to two PES local minima with a total energy difference of 0.21 eV, and exhibit a significant difference in their optical properties, with a maximum achievable theoretical shift of the main absorption peak of 20 nm (Figure 4b). The mechanically tuneable electrostatic interaction so provides the P(VDF-TrFe)/LD700 system with so-called mechanophore units making the nanowire emission sensitive to tensile stress. Instead, for the P(VDF-TrFe)/pyrromethene system no significant changes of the optical properties can be expected upon applying the mechanical stress, because different P(VDF-TrFe) arrangement do not modify the electronic distribution of the light-emitting molecule. The DFT-optimized supramolecular organization of the P(VDF-TrFe)/pyrromethene complex is shown in Figure S13. For this ground-state conformation, a lowest singlet excited state ($S_1$, with the strongest oscillator strength) is predicted at 3.04 eV. This is almost unmodified by an external stress, leading to a relatively rigid spatial shift between the pyrromethene and the polymer along its main chain axis (Figure S14).

The here presented mechanisms involve a soft mechanochemical actuation, namely the influence of the directional, tensile stress on weak, electrostatic forces mediated by the coupled counter-ion in the almost periodic and anisotropic structure of the P(VDF-TrFe) nanowires. Largely studied as tool to selectively form and break covalent bonds,[15,16,27] mechanophores making optical properties sensitive to strain or stress can also rely on electrostatic coupling. In this respect, stress and deformation act as stimuli to redirect optically-active units, and make their localization in a specific PSE minimum more likely, thus affecting the nanowire emission properties.

The investigated correlation between a stress and the emission properties of the LD700/ClO$_4^-$ coupled system suggests an important role played by the mechanical properties of the nanowire array at macroscale. For instance, with stiffer arrays, comparable levels of applied stress can entail lower values of the associated strain and consequently, a reduced spatial shift between the polymer chains and the LD700/counter-ion. Also, stiffer arrays might increase the stability of mechanophore units in the involved PES local minima. Tuning of the mechanical properties can be achieved through properly addressing the electrospinning parameters to increase the number of the inter-fibers joints[28] or to tailor the overall porosity in nanocomposites.[29]

In summary, the effect of tensile bending strain on the optical properties of dye-doped nanocomposite wires made of a PVDF copolymer is significant. Results indicate a red-shift up to the scale of 5 nm in counter-ion dye-doped piezoelectric wires. Consistent with the results of first-principles DFT simulations, this effect can be rationalized with the formation of mechanically-tunable electrostatic interaction, namely mechanophore units making nanowire emission sensitive to tensile stress. These electrostatic mechanophores might provide anisotropic, hybrid molecular systems with novel functionalities, including mechanochromisms and the consequent optical sensing capability to monitor local mechanical stress and damage also in transparent samples and soft robots.

*Experimental Section*

*Nanowires fabrication and characterization.* P(VDF-TrFE) (75/25 weight%, Solvay Solexis) was dissolved in a 3:2 volume ratio of dimethylformamide/acetone (DMF/acetone, Sigma Aldrich) at a polymer/solvent concentration of 21% w/w. PMMA (120K, Sigma Aldrich) was dissolved in chloroform at a polymer/solvent concentration of 25% w/w. LD700, pyrromethene and DCM were added to solutions with relative concentration 1-1.5% w/w with respect to the polymers. Electrospinning was performed by placing 1-1.5 mL of solution into syringes tipped with a 21-27-gauge stainless steel needles. The positive lead from a high voltage supply (EL60R0.6-22, Glassman High Voltage) was connected to the metal needle for application of bias values of 15-25 kV. The solution was injected into the needle at a constant rate of 1 mL/hr with a syringe pump (33 Dual Syringe Pump, Harvard Apparatus). A cylindrical collector was placed at a distance of 6 cm from the needle and negatively biased for the fabrication of aligned arrays. All the fabrication steps were performed at room temperature with air humidity of about 40%. The morphological analysis was performed by a Nova NanoSEM 450 system (FEI), using an acceleration voltage around 5 kV and an aperture size of 30 mm. The mechanical properties were investigated by dynamic mechanical analysis (DMA Q800, TA Instruments, New Castle, DE), in tensile mode. Specimen dimensions were approximately 6.5×10.0 mm (width×length) with thicknesses between 20 and 30 μm. FTIR measurements were carried out by a Spectrum 100 system (Perkin-Elmer Inc.), equipped with an IR grid polarizer (Specac Limited, UK), consisting of 0.12 mm wide strips of aluminium. The beam, incident orthogonally to the plane of the sample, was polarized alternatively parallel or orthogonal to the main axis of fiber alignment.

*Mechano-tunable emission.* Two synchronized linear stages (Physik Instrumente GmbH & Co. KG, Karlsruhe, Germany) were used to apply cyclic dynamic bending stress. A fast camera (FAST CAM APX RS Photron) was used to collect videos (1024×1024 pixel$^2$) at 250 fps, corresponding to acquisition times of 4 ms. The light emitted by the nanowires was collected synchronously with their mechanical movement, exciting samples by a diode laser with wavelength $\lambda_{exc}$=405 nm and collecting continuous sequences of 26 photoluminescence spectra by a fiber-coupled monochromator equipped with a charged coupled device (Figure S8) and careful calibration procedures on samples at rest were carried out to take into account contributions related to geometry-related photon collection efficiency and angle-modulated photon scattering (full details in Supporting Information).

*Computational Details.* The P(VDF-TrFE) copolymer was investigated using the plane-wave Quantum Espresso software package[30] using the Perdew-Burke-Ernzerhof functional and ultrasoft pseudopotential with an energy (density) cutoff of 80 (800) Ry. A Gaussian smearing of 0.001 Ry was used to enhance convergence and the threshold for energy convergence was set at $10^{-6}$ Ry. We used a cell of about 29 Å in the vertical direction and a dipole correction scheme,[31] while in the lateral direction the distance was about 24 Å. The simulated cell included 12 atoms and 9.1 million G-vectors. The geometry was fully relaxed. Calculations for LD700 with $ClO_4^-$ and for pyrromethene were performed with TURBOMOLE program package[32] using the PBE0 functional together with empirical dispersion[33] and the def2-TZVP basis set for optimization.[34] The TD-DFT absorption spectra were then computed with the same parameters but with the def2-TZVPD basis set.[35]

*Acknowledgements*

The research leading to these results has received funding from the European Research Council under the European Union's Seventh Framework Programme (FP/2007-2013)/ERC Grant Agreement n. 306357 (ERC Starting Grant "NANO-JETS"). The Apulia Network of Public Research Laboratories WAFITECH (09) is also acknowledged. Maria Moffa is acknowledged for her support with mechanical measurements.

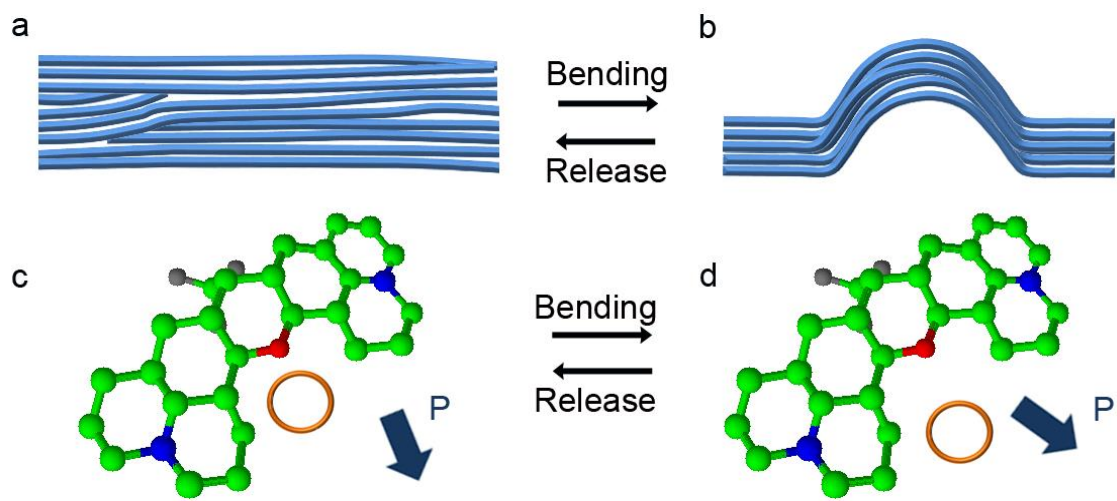

**Figure 1.** Schematics of free-standing arrays of nanofibers before (a) and during (b) bending. c,d: Molecular structure of LD700, and scheme of the counter-ion shift process leading to mechano-tunable emission. The counter-ion, $ClO_4^-$ (orange ring) leads to a different configuration of the system during nanofiber bending, namely to reaching a different local minima of the potential energy surface. **P** indicates a piezo-dipole, roughly perpendicular to the nanofiber length. Colours in the molecular structures indicate different atoms. Red: oxygen, grey: fluorine, blue: nitrogen, green: carbon.

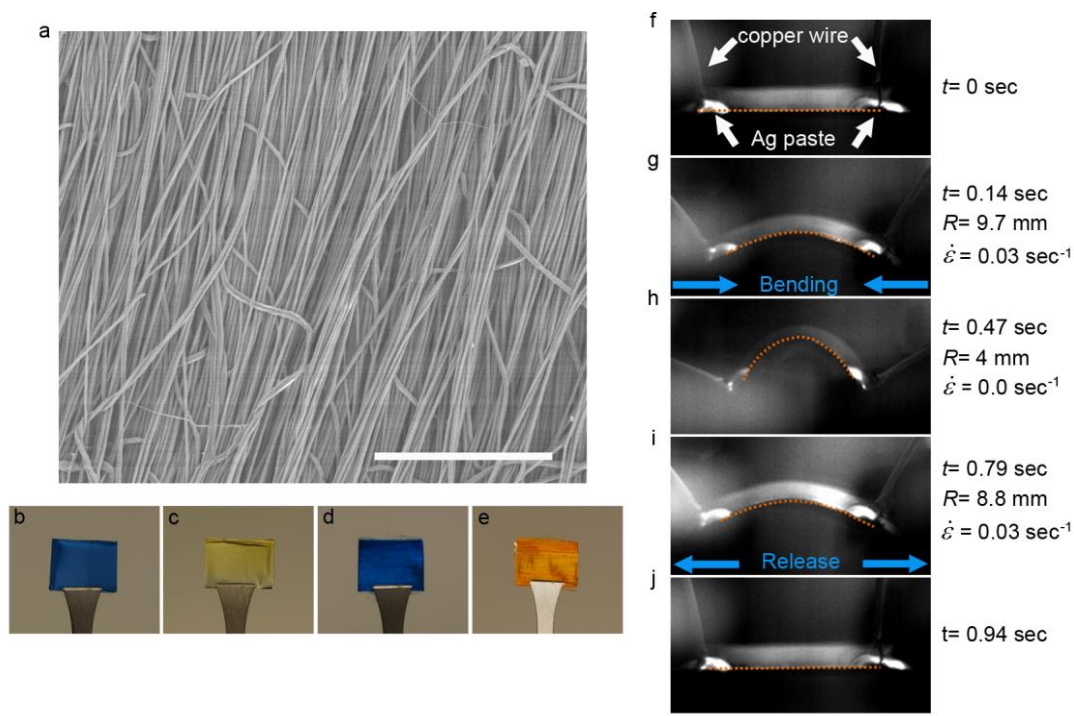

**Figure 2.** (a) SEM micrograph of the composite nanowires based on P(VDF-TrFe) doped with LD700. Scale bar: 10 µm. (b-e) Photographs of the arrays of composite nanowires. From left to right: P(VDF-TrFe)/LD700, P(VDF-TrFe)/pyrromethene, PMMA/LD700 and PMMA/DCM, respectively. (f-j) Sequence of frames captured during bending of the nanowires. $R$ and $\dot{\varepsilon}$ are the bending radius and strain rate at each time, respectively.

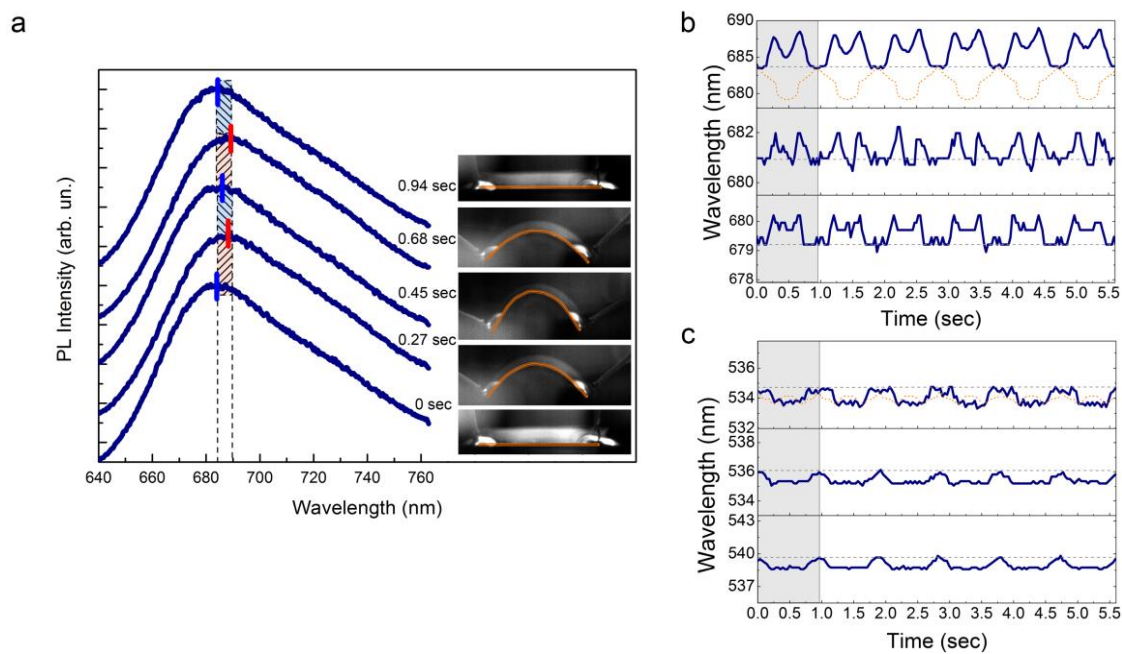

**Figure 3.** (a) PL spectra from arrays of P(VDF-TrFe)/LD700 fibers at different times during bending. Blue and red markers on the spectra highlight the position of the PL peak at each time, while photographs capture the corresponding mechanics. Spectra are vertically shifted for better clarity. (b, c) Plots of the emission peak wavelength vs time for P(VDF-TrFe)/LD700 fibers (b), and P(VDF-TrFe)/pyrromethene (c) fibers respectively. Emission temporal profiles are shown for three different samples for each type of fiber. The shadow area indicates the first bending interval, and dashed horizontal lines highlight the initial wavelength peak for unperturbed samples. Finally, the dotted profiles in the top panel for each nanowire type are the calculated contributions to shifts coming from optical detection (namely, for the different distance of the sample and of the collection lens, under static conditions).

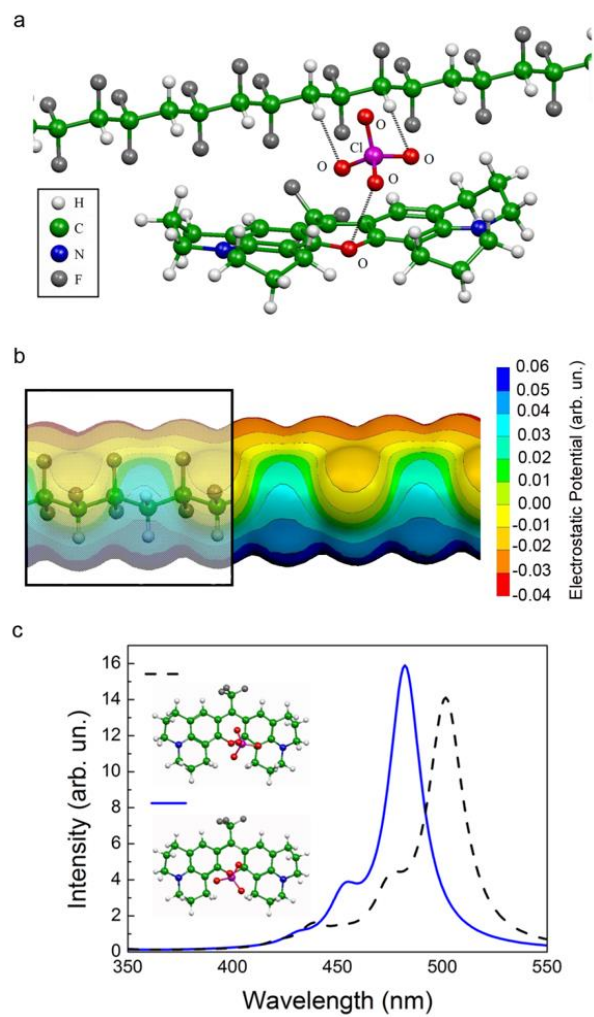

**Figure 4.** (a) Optimized conformation for the P(VDF-TrFe)/LD700 system, mediated by the $ClO_4^-$ counter-ion. Dotted lines represent the shortest distances to $ClO_4^-$. (b) Colour map of the electrostatic potential of the P(VDF-TrFe) chain. (c) TD-DFT absorption spectra for two different conformations of LD700/$ClO_4^-$, shown in the inset.

# Supporting Information

**Electrostatic mechanophores in tuneable light-emitting piezo-polymer nanowires**

*Luana Persano\*, Andrea Camposeo, Alexandrs V. Terentjevs, Fabio Della Sala\*, Eduardo Fabiano, Martina Montinaro, Dario Pisignano\**

Dr. L. Persano, Dr. A. Camposeo, Prof. D. Pisignano
NEST, Istituto Nanoscienze-CNR, Piazza S. Silvestro 12, I-56127 Pisa, Italy
E-mail: luana.persano@nano.cnr.it
Dr. A. Camposeo, Dr. F. Della Sala, Dr. E. Fabiano, Prof. D. Pisignano
Center for Biomolecular Nanotechnologies @UNILE, Istituto Italiano di Tecnologia, via Barsanti, I-73010 Arnesano (LE), Italy
Dr. A. V. Terentjevs
Istituto Nanoscienze-CNR, Via per Arnesano, I-73100, Lecce, Italy
F. Della Sala, E. Fabiano
Institute for Microelectronics and Microsystems (CNR-IMM), Via Monteroni, Campus Unisalento, I-73100 Lecce, Italy
E-mail: fabio.dellasala@le.imm.cnr.it
M. Montinaro, D. Pisignano
Dipartimento di Matematica e Fisica "Ennio De Giorgi", Università del Salento, via Arnesano, I-73100, Lecce, Italy
E-mail: dario.pisignano@unisalento.it

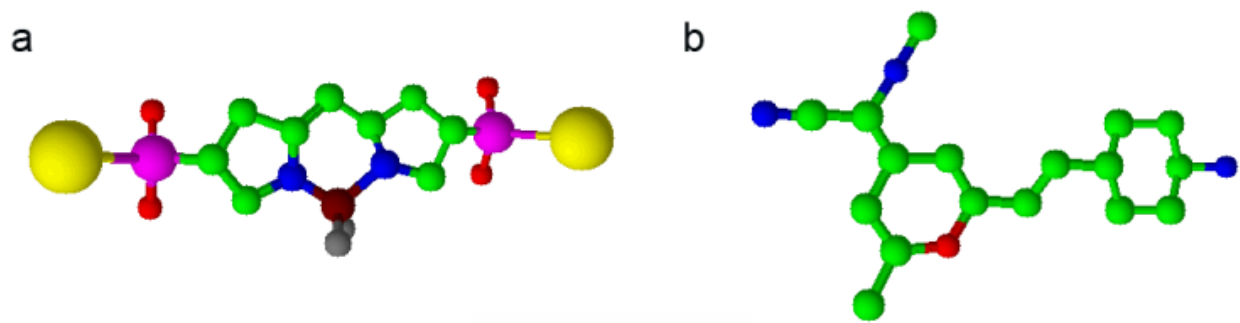

Figure S1. Molecular structure of pyrromethene (a) and [2-[2-[4-(dimethylamino)phenyl]ethenyl]-6-methyl-4H- pyran-4-ylidene]-propanedinitrile (DCM, b). Colours indicate different atoms. Yellow: sodium, red: oxygen, magenta: sulfur, green: carbon, blue: nitrogen, maroon: boron, grey: fluorine.

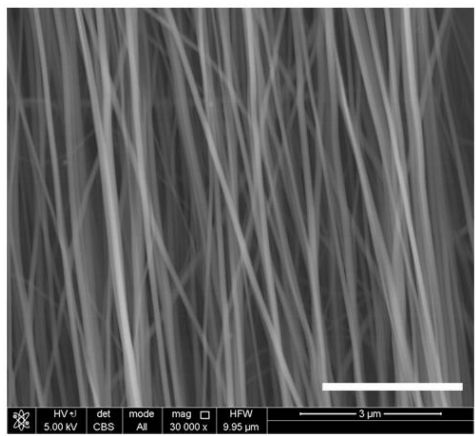

Figure S2. SEM micrograph of an uniaxially-aligned array of composite nanowires made of P(VDF-TrFe) with pyrromethene (scale bar: 3 μm).

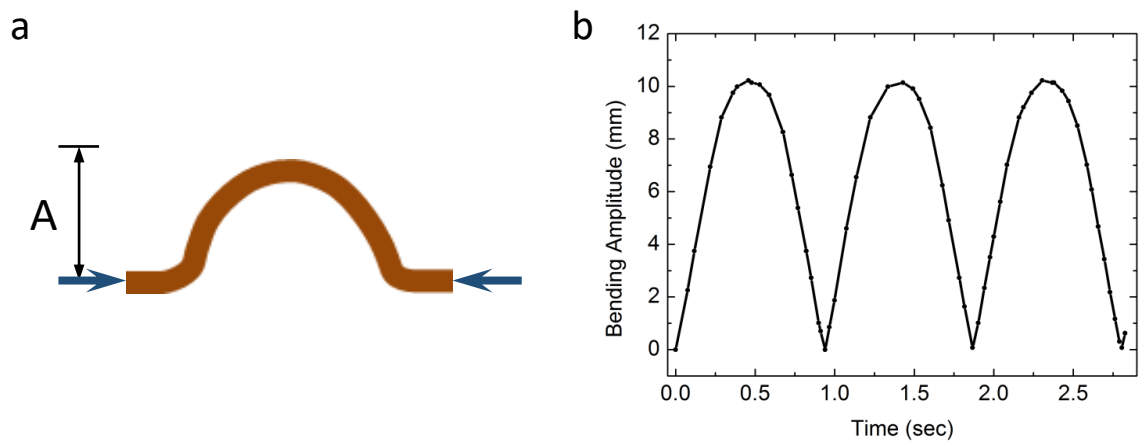

Figure S3. a: Schematic illustration of the bending movement. *A* is the out-of-plane bending amplitude. b: Plot of the bending amplitude vs time. The bending frequency is about 1 Hz and each half-period is well described by a uniformly accelerated motion with initial speed $v_0$= 44 mm/s and acceleration, $a$= 92 mm/s$^2$.

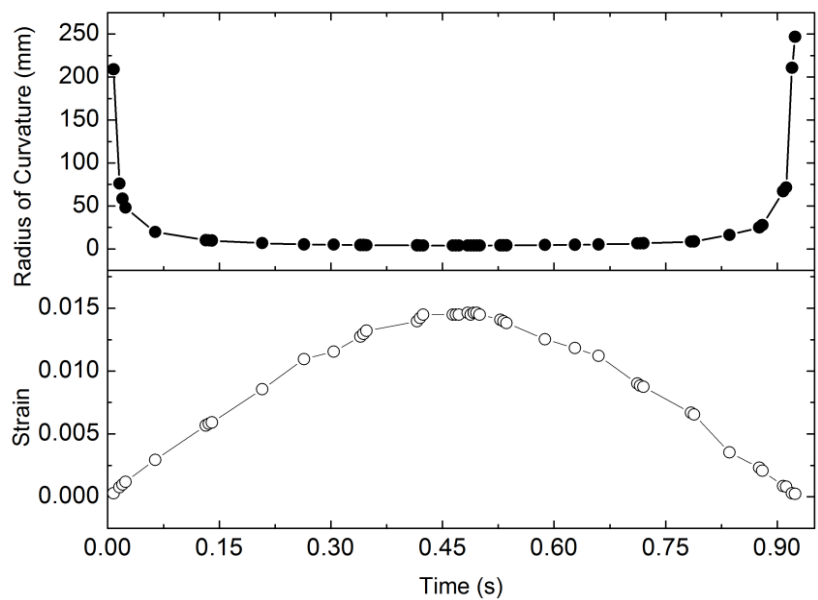

Figure S4. Plot of the radius of curvature $R$ (a) and maximum local strain values (b) vs time during bending. The maximum strain at the outer surface of the fibers is obtained as $\varepsilon_m = c/R$, where $c$ is the distance of the top surface of the fibers from the neutral strain position,[S1] and $R$ is the radius of curvature.

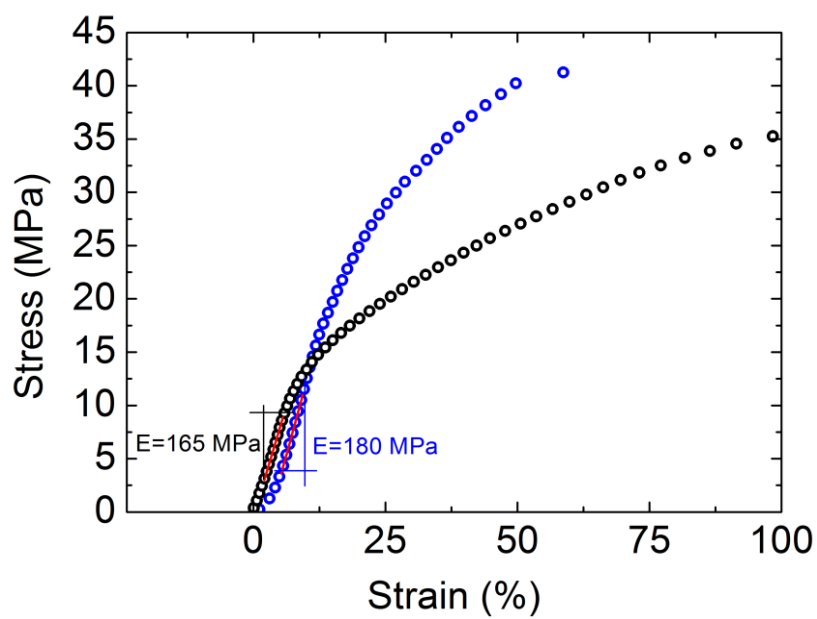

Figure S5. Stress-strain curves measured for aligned arrays of fibers of P(VDF-TrFe) (black dots) and

P(VDF-TrFe)/LD700 (blue dots). Red lines are fits in the linear regions highlighted in the figure, and

*E* are the corresponding values of the Young Modulus.

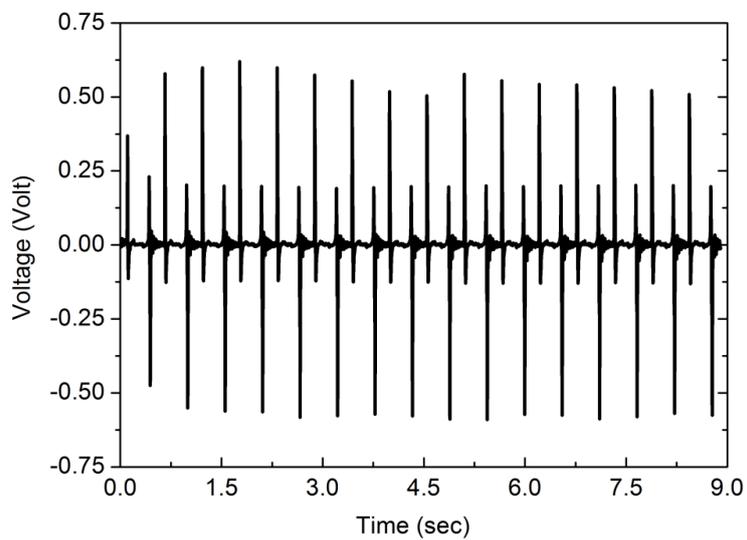

Figure S6. Measured voltage response of an array of P(VDF-TrFe) fibres doped with LD700, under cycling bending at about 2 Hz.

**Optical measurements**

Steady state photoluminescence (PL, Figure S7) spectra were obtained by exciting the sample with a diode laser and measuring the spectral features of the emission by a spectrometer (USB 4000, Ocean Optics). The angular dependence of the PL was obtained by collecting the emitted photons by an optical fiber mounted on a rotating stage. The angular resolution of the measurements was 0.3°.

The experimental system used to collect the light emitted by the samples, synchronously to their mechanical movement, is schematized in Figure S8. The PL was excited by a diode laser with wavelength $\lambda_{exc}$=405 nm and maximum emission power of 5 mW. The latter was attenuated by neutral density filters to achieve an incident power of 0.2 mW onto the sample surface. This value guaranteed a good photostability of the emission within a typical measurement time, namely a variation of the PL intensity <10 % and a shift of the peak wavelength < 0.5 nm on a time interval of the order of minutes. Such set of parameters of the excitation laser provided also a good signal-to-noise ratio (>20) for those emission spectra acquired with the shortest integration time (20 ms).

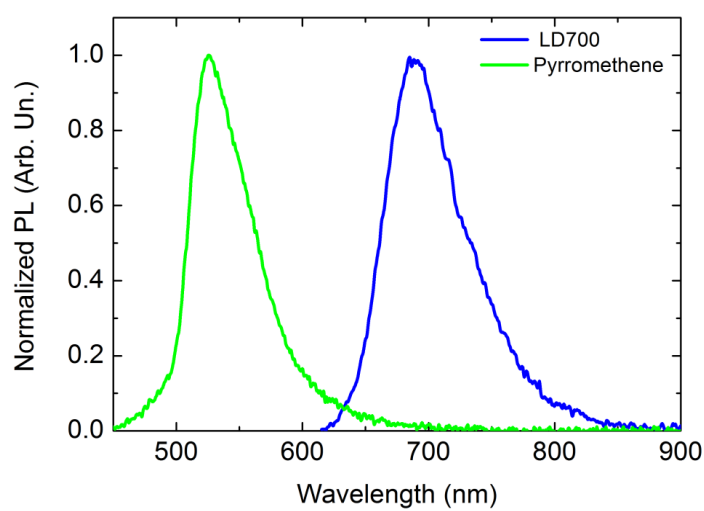

Figure S7. Normalized photoluminescence (PL) spectra of an array of P(VDF-TrFe) electrospun fibres doped with LD700 (blue line) and with pyrromethene (green line).

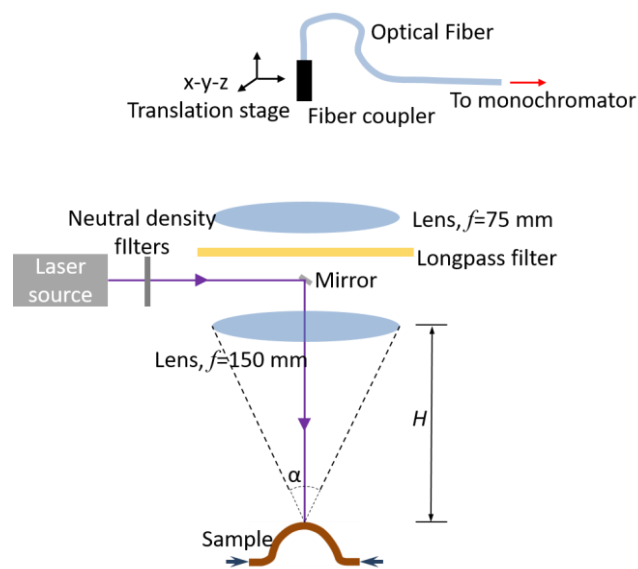

Figure S8. Scheme of the optical set-up used to collect the light emitted by the nanowires during mechanical bending. The excitation laser is focused in the center of the sample through a lens with $f$=150 mm. The light collection is synchronous with mechanical movement.

The excitation laser spot size onto the samples surface was about 200 μm and was precisely aligned in order to keep it fixed on the sample during the mechanical bending. In addition, within the excited area, we estimated that the maximum height difference between points along the profile of curvature induced by the mechanics of bending is much smaller than the spot size (by one order of magnitude), thus not affecting the overall intensity of the emitted light. This allows us to excite optically the same area during symmetric mechanical bending. The sample PL was collected by a lens (focal length, $f$=150 mm) in a back-scattering configuration, and coupled to an optical fiber by a second lens ($f$=75 mm). The optical fiber coupler was mounted on a precision, micrometric, three-axes translation stage, used to maximize the light coupled into the optical fiber. A longpass filter, with cut-on wavelength at 450 nm, was positioned between the two lenses of the collecting system to remove spurious light coming from the excitation laser beam, diffused by the sample surface. The PL spectra were measured by a monochromator (iHR320, Jobin Yvon), equipped with a charged coupled device (CCD) detector (Symphony, Jobin Yvon). In a typical time-resolved optical measurement during the symmetric mechanical bending, the sample was continuously excited, while a sequence of 26 PL spectra were acquired for each bending cycle, with 20 ms integration time. The time interval between two consecutive acquired spectra was 36 ms. Minor temporal fluctuations of the emission peak are attributable to variations of the collection conditions due to local photon scattering phenomena during bending.

The excitation and collection systems were fixed mechanically to the optical table and PL spectra are acquired avoiding any spurious effect related to movement (Figure S9). During dynamic bending measurements, the distance, $H$, between the sample excited area and the collecting lens varied in an interval 14.5 cm ($H_{min}$) - 15.5 cm ($H_{max}$). As a consequence, the collection angle, α, varied in a range 18° ($α_{min}$) - 20° ($α_{max}$) (Figure S8).

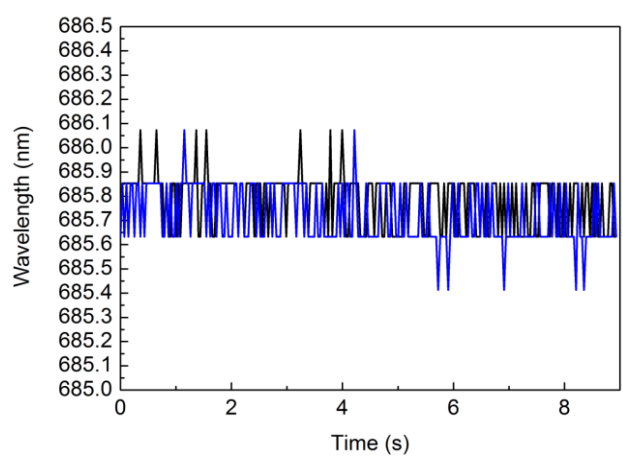

Figure S9. Position over time of the PL peak wavelength of P(VDF-TrFe)/LD700. Black line: both sample and translation stages fixed. Blue line: sample fixed, translation stages in motion.

The detection system was designed to minimize as much as possible the variation of the collection efficiency of the emitted photons during the mechanical movement of the samples, however light scattering by the nanofibrous samples, self-waveguiding of the emitted photons along the fibers and self-absorption effects might introduce small shifts of the PL bands. This is suggested by the measured angular dependence of the PL peak wavelength in an angle range of about ±10° (Figure S10), highlighting a spectral shift of 1.8 nm in P(VDF-TrFe)/LD700 samples. In order to account for these contributions, we implemented a calibration procedure for each sample, by measuring the spectral shift under static conditions.

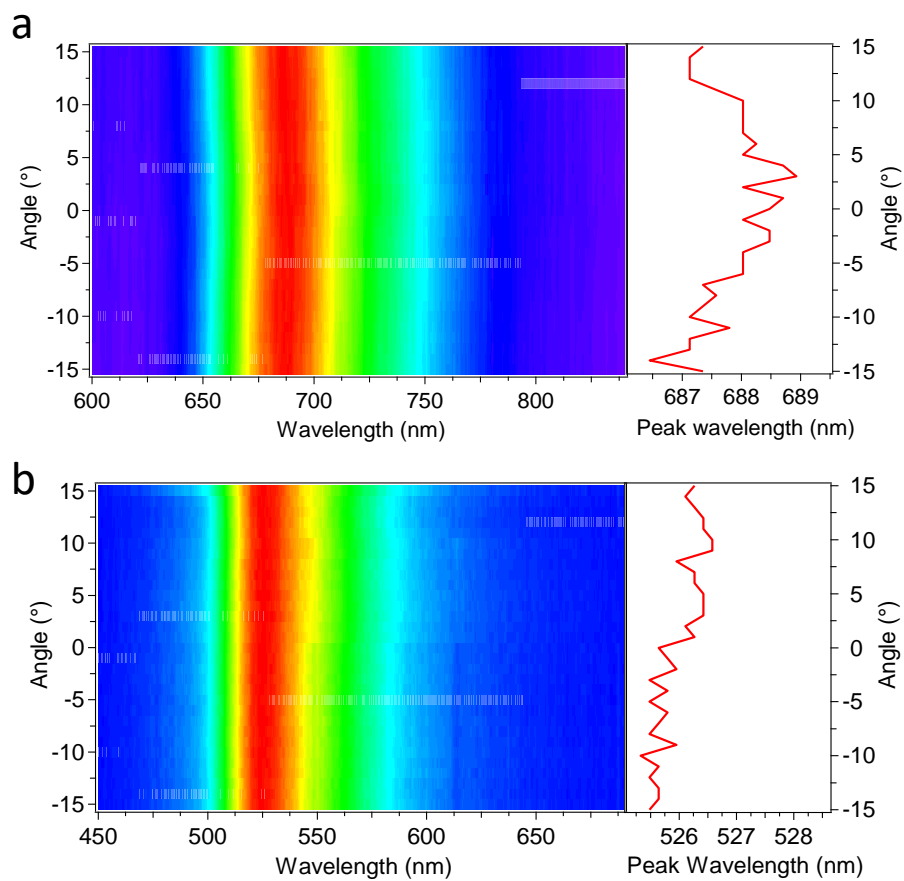

Figure S10. Angular dependence of the PL spectra (left panels) and of the PL peak wavelength (right panels) for P(VDF-TrFe) fibers with LD700 (a) and with pyrromethene (b).

To this aim, calibration measurements were performed for each sample. PL spectra were acquired upon varying the relative distance, $H$, between the excited spot and the collecting lens in the range $H_{min}$-$H_{max}$, while maintaining the samples in a flat configuration. By measuring the spectral shifts for each sample during bending, $\lambda_b(H)$, and the time dependence of the relative distance between the excited spot and the collecting lens in a typical bending cycle, $H(t)=H_{max}-A(t)$, where $A(t)$ is the bending amplitude during the mechanical movement, we could calculate the expected spectral shift during movement due to the detection system [$\lambda_b(t)$, dotted profiles in Figure 3b,c]. In this way, the implemented calibration procedure allows the shift due to the detection system to be precisely determined and directly compared to those measured in dynamic experiments.

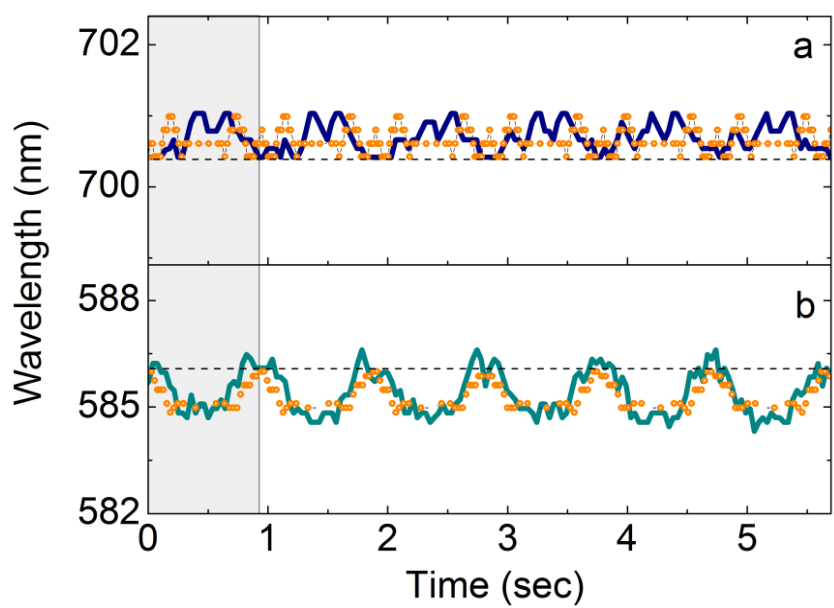

Figure S11. Plots of the emission peak wavelength vs time for PMMA/LD700 fibers (a), and PMMA/DCM (c) fibers, respectively. The shadow area indicates the first bending interval, and dashed horizontal lines highlight the initial wavelength peak for unperturbed samples. Finally, the dotted profiles are the calculated contributions to shifts coming from optical detection (analogous to Fig. 3b,c).

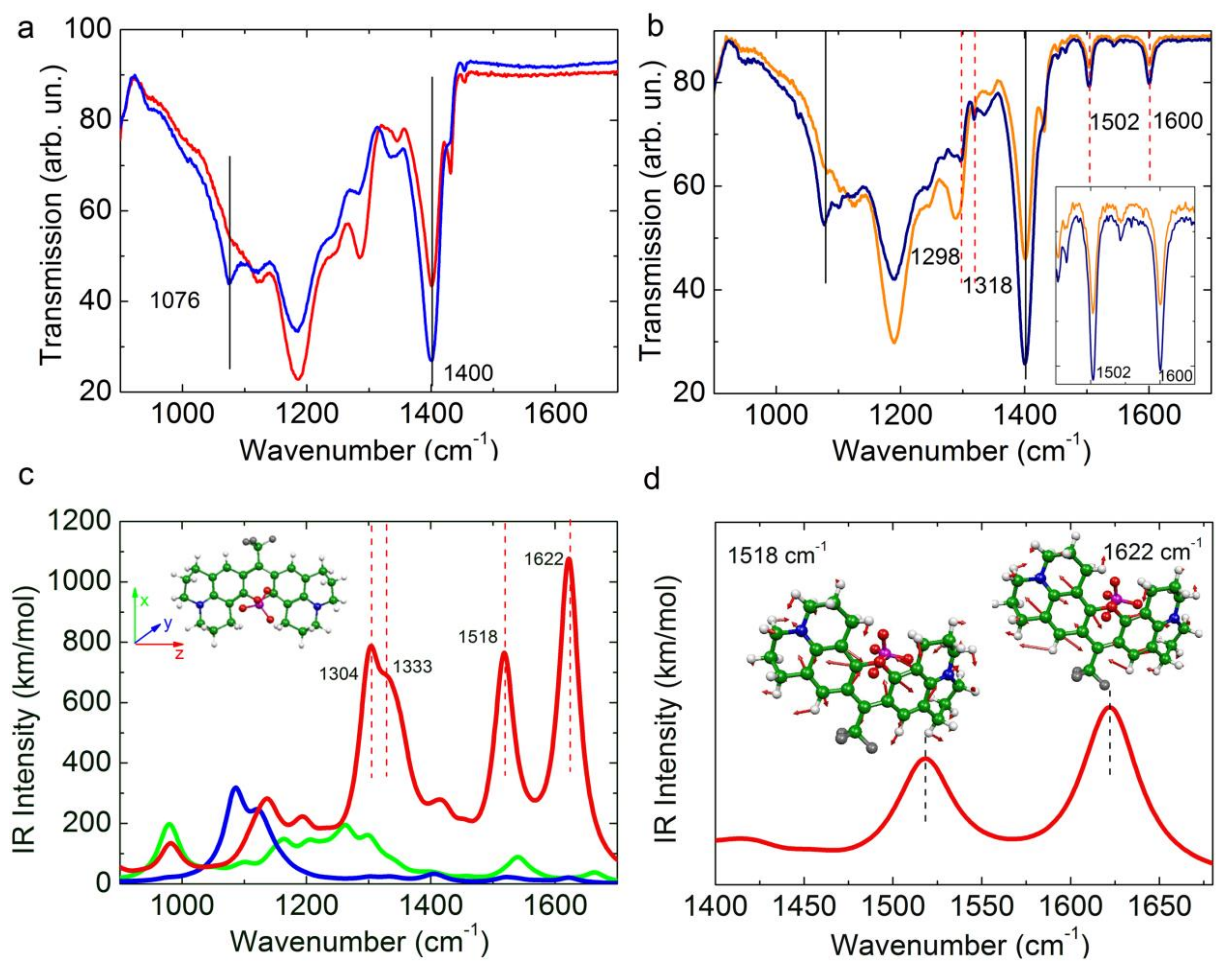

Figure S12. FTIR spectra measured under different incident beam polarizations for aligned fibers of P(VDF-TrFe) (a) and P(VDF-TrFe)/LD700 (b). Light is polarized either parallel [blue lines in (a) and (b)] or perpendicular (red line in (a) and orange line in (b)) to the long axis of the fibers. Vertical black lines mark bands associated with vibrations that depend on chain orientation (1076 and 1400 cm⁻¹), highlighting a significant increase in the intensity of these bands for light polarized along the longitudinal axis of the fiber (blue lines). Dashed red lines in (b) marks peaks associated with the dye molecule. The inset in (b) is a magnified view, highlighting the increase of the intensity of peak at 1502 and 1600 cm⁻¹ for light polarized along the axis of the fibers (blue line). (c) Computed (DFT/PBE) IR absorption spectra of LD700 with light polarized along the cartesian axis, $z$ (red line), $x$ (green line), or $y$ (blue line), according to the directions shown in the inset. Vertical, dashed red lines highlight a significant increase of peaks polarized along the long molecular axis ($z$). (d) Magnified view of the high energy region of the computed IR spectrum for light polarized along $z$, displaying the involved atoms oscillations (normal modes) for the peaks at 1518 cm⁻¹ and 1622 cm⁻¹ (arrows on the molecular structures). Colours in the molecular structures indicate different atoms. Red: O, grey: F, blue: N, green: C, purple: Cl, white: H.

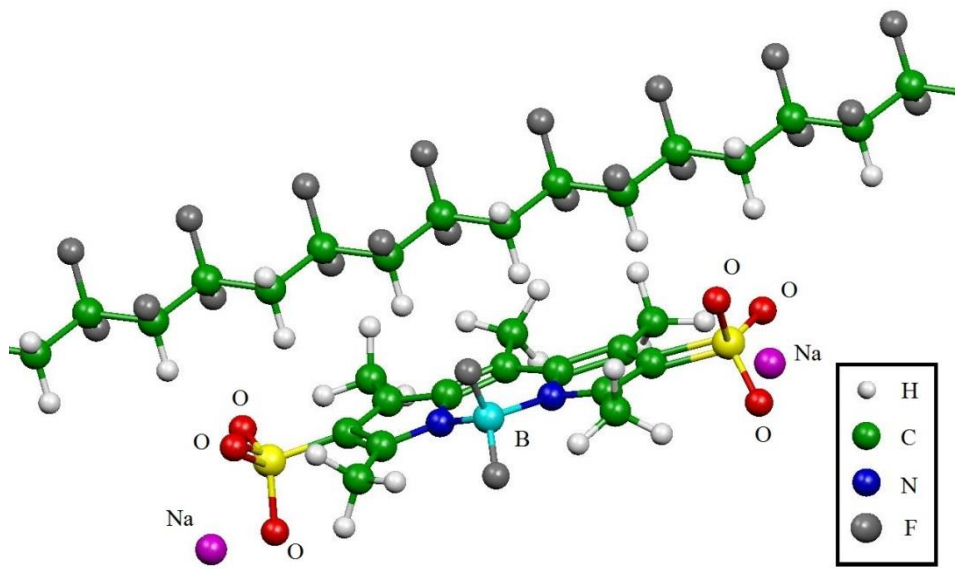

Figure S13. Optimized structure for the P(VDF-TrFe)/pyrromethene system.

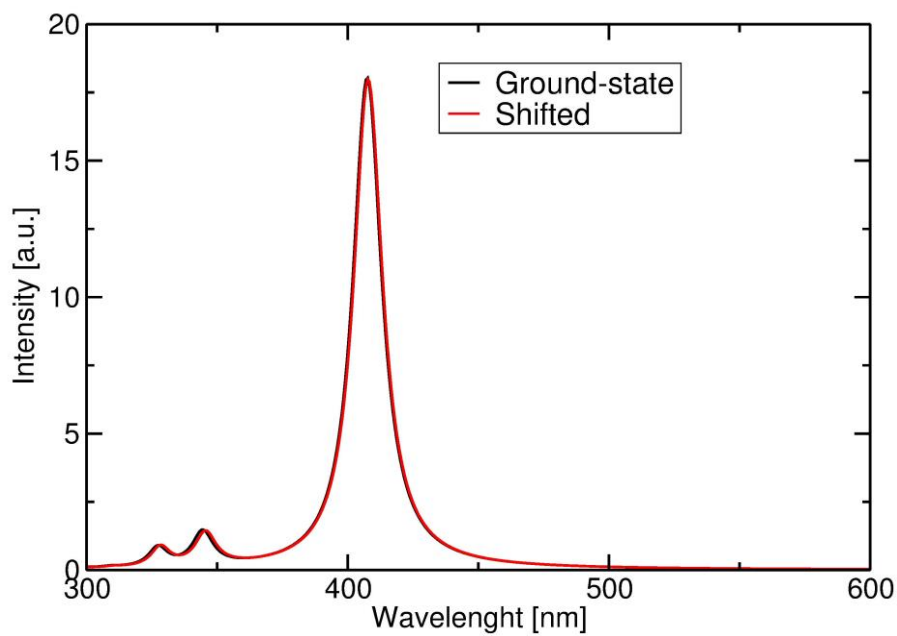

Figure S14. TD-DFT absorption spectra of the P(VDF-TrFe)/pyrromethene system in the ground-state conformation (see Fig. S13) and when the polymer is shifted along its main axis by 1.3 Å